# Mechanistic Insights into the Challenges of Cycling a Non-aqueous Na-O$_2$ Battery

Tao Liu[1], Gunwoo Kim[1,2], Mike T. L. Casford[1], Clare P. Grey[1]*

[1]*Chemistry Department, Lensfield Road, Cambridge, UK CB2 1EW*

[2]*Cambridge Graphene Centre, University of Cambridge, Cambridge, UK CB3 0FA*

**Abstract**

Superoxide-based non-aqueous metal oxygen batteries have received considerable research attention, as they exhibit high energy densities and round-trip efficiencies. The cycling performance, however, is still poor. Here we study the cycling characteristic of a Na-O$_2$ battery using solid-state nuclear magnetic resonance, Raman spectroscopy and scanning electron microscopy. We found that the poor cycling performance is primarily caused by the considerable side reactions stemming from the chemical aggressiveness of NaO$_2$ both as a solid phase and dissolved species in the electrolyte. The side reaction products cover electrode surfaces and hinder electron transfer across the electrode-electrolyte interface, being a major reason for cell failure. In addition, the available electrode surface and porosity change considerably during cell discharging and charging, affecting the diffusion of soluble species (superoxide and water) and resulting in inhomogeneous reactions across the electrode. This study provides insights into the challenges associated with achieving long-lived superoxide based metal-O$_2$ batteries.

Graphical abstract

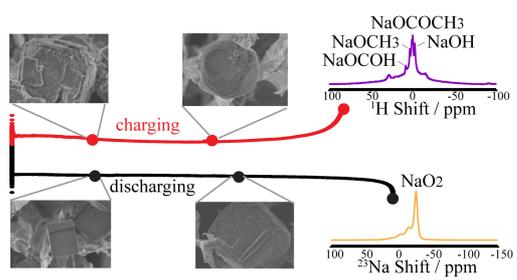

Rechargeable non-aqueous alkali metal (Li, Na, or K)-$O_2$ batteries are promising candidates for next-generation energy storage because they possess significantly higher theoretical energy densities than the state of the art lithium ion batteries.[1-3] Their operation on discharge commonly involves the reduction of $O_2$ to precipitate a solid phase product in the porous cathode; on charging, the solid product is decomposed releasing $O_2$. The type of discharge products, on the other hand, can differ among the three $O_2$ batteries. Superoxide,[2-4] peroxide[5-7] and hydroxide phases[8-9] of alkali metals have been reported so far and the phase via which the battery cycles, has a decisive impact on the battery performance (energy density, rechargeability, cycling lifetime, round-trip efficiency). Currently, there is an increased interest in superoxide-based alkali metal-$O_2$ batteries, because they typically exhibit much smaller discharge-charge overpotentials (as low as 0.2 V) than batteries based on the other two phases, mitigating many problems associated with high overpotentials. Nevertheless, superoxide-based batteries generally show poor cycling performance. Understanding the cycling characteristics and the failure mechanism of superoxide-based batteries is therefore essential to further improve their performance, and may also help understand the peroxide and hydroxide chemistries. In this work, we monitor the reversible and irreversible processes in Na-$O_2$ batteries using microscopic and spectroscopic techniques, to provide mechanistic insights into the challenges associated with cycling a superoxide-based battery.

A Na-$O_2$ battery was constructed using a sodium metal anode, a porous reduced graphene oxide (rGO) cathode and 0.25 M NaClO$_4$/dimethoxyethane (DME) as the electrolyte, as described in the Experimental Section. The cell was discharged and charged in 1 bar pure O$_2$, typically in a voltage window of 1.8-2.8 V versus Na/Na$^+$. The discharge-charge profile of the 1$^{st}$ cycle and the cycling performance of the Na-O$_2$ battery are shown in Fig. 1(a) and (b) respectively. Very flat discharge and charge plateaus are observed at 2.1 and 2.45 V respectively, with a voltage gap of only ~0.35 V consistent with previous reports.[10] The cell shows a capacity of 6 mAh, which corresponds to 40,000 mAh/g$_{carbon}$ or 12 mAh/cm$^2$, much larger than that of a mesoporous carbon electrode (typically < 2 mAh, 10,000 mAh/g$_{carbon}$ or 3 mAh/cm$^2$);[11] this observation is attributed to the hierarchically macroporous framework in the rGO electrode that allows dense and continuous growth of the cubic discharge product beyond 10 μm in size (Fig. S1), whereas in a mesoporous electrode the size of the discharge product is smaller and pore clogging is more likely to occur causing an early finish of the discharge. This idea of promoting the growth of large discharge product crystals in a macroporous electrode also applies to Li-O$_2$ batteries.[8] In a DME-based electrolyte, the Li$_2$O$_2$ toroidal particles formed on discharge can also grow beyond 10 μm, leading to a high capacity of up to 8 mAh or 16 mAh/cm$^2$ (Fig. S1).

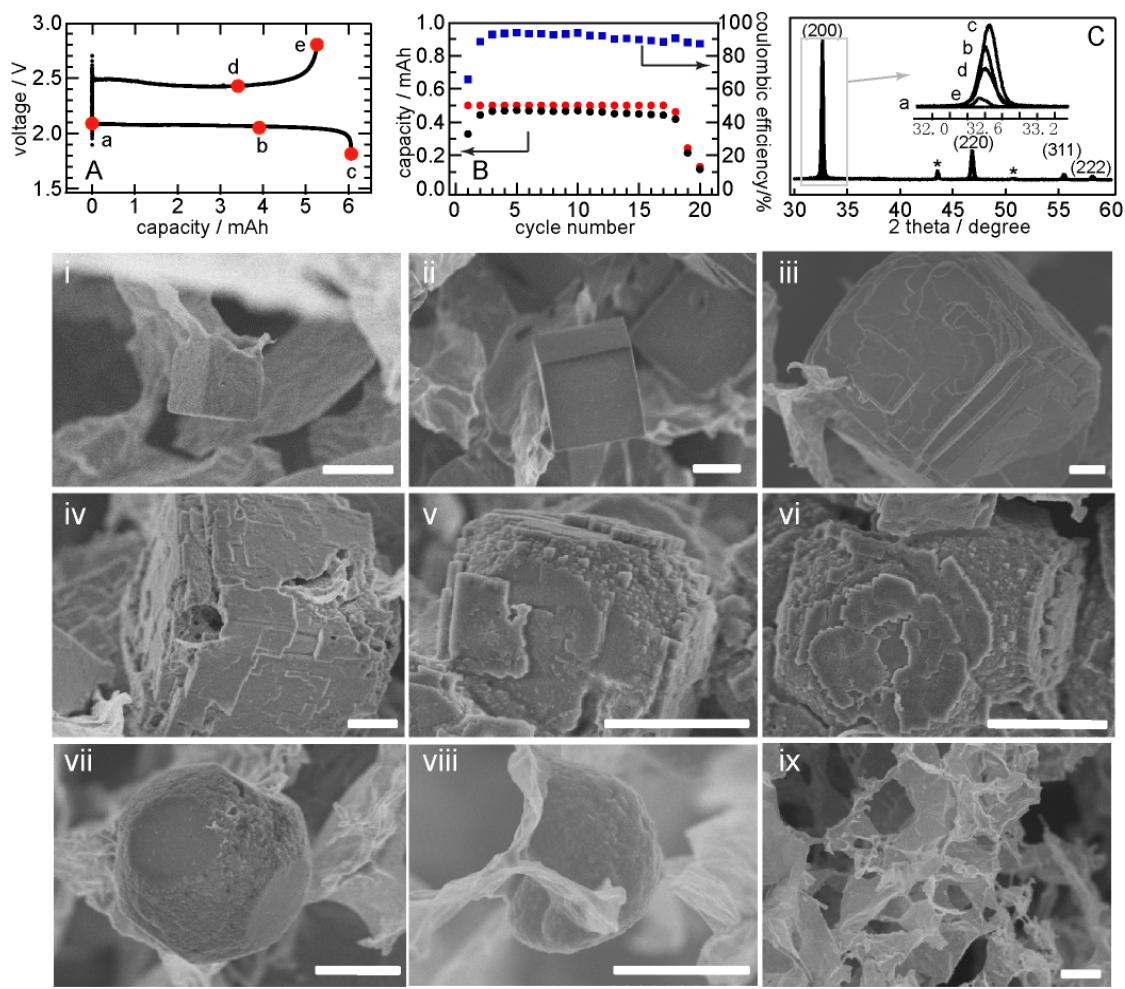

**Figure 1**. Electrochemistry of Na-$O_2$ batteries, and PXRD and SEM characterization of the rGO electrodes from cells finishing at different states of charge during the 1$^{st}$ cycle. A typical discharge-charge curve (A) of a Na-$O_2$ cell made using an rGO electrode and 0.25 M NaClO$_4$/DME electrolyte; a 50 μA (0.1 mA/cm$^2$) current was used; cycling performance of a cell (0.5 mAh per discharge) is shown in (B) and blue square, red and black circles denote coulombic efficiency, discharge and charge capacities, respectively. PXRD patterns were measured on rGO electrodes from cells terminated at the states labelled in (A), and show that NaO$_2$ is the only crystalline phase formed and that it is removed on charge. SEM shows that the NaO$_2$ crystals grow larger as discharge proceeds (i-iii); (iv-ix, from charged electrodes) in sequence show how the NaO$_2$ crystals are etched away on charging. Images (i-iii) were taken from electrodes with 0.5, 2 and 6 mAh discharge capacities; images (iv-vi) were taken from the same charged electrode following a discharge of 6 mAh and then a charge of 2 mAh, (vii) from that with a charge of 4 mAh, and (viii-ix) from the electrode charged to 2.8 V. All scale bars represent 2 μm.

To verify the chemical nature of the discharge product, powder X-ray diffraction (PXRD) measurements (Fig. 1C) were performed at various points of the discharge-charge curve (Fig. 1A). They confirm that sodium superoxide ($NaO_2$) is the dominant crystalline phase formed on discharge and it is removed on charge. Notably, this removal is not complete (Fig. 1C), consistent with the inability to fully recharge the battery (Fig. 1A). As the $NaO_2$ crystals grow larger, all visible diffraction peaks are shifted to higher angles (Fig. 1C, inset), which corresponds to a reduced unit cell in larger crystals and suggests that $NaO_2$ may occlude ion/molecules in the crystals during cycling. Previously, crystalline phases of sodium peroxide ($Na_2O_2$)[7] and its hydrate ($Na_2O_2 \cdot H_2O$)[12-16] were also reported to form on discharge in Na-$O_2$ cells with an ether electrolyte and they decompose at around 4.0 and 3.0 V respectively. We have charged our batteries up to 4.0 V (Fig. S2) and little capacity was observed at those voltages, indicating that sodium peroxide chemistry is negligible in our cells.

Further investigation of the cell separators using scanning electron microscopy (SEM) shows that sodium superoxide particles also grow on the insulating glass fibres and they accumulate with more cycles (Fig. S3). This phenomenon is consistent with the recent finding that both the formation and decomposition of $NaO_2$ involves a solution mechanism in the presence of a trace amount of water:[10,17-19] e.g., on charging, with the aid of water $NaO_2$ dissolves from the bulk crystal surfaces and then the superoxide species diffuses into the electrolyte before decomposing at the electrode surface. The small amount of water in our cells is confirmed to come from the water impurities from the $O_2$ purge line and estimated to be ~50 ppm by a Karl-Fischer measurement after a purge.

To study the solution process, we have monitored the formation and decomposition of $NaO_2$ crystals at various states of discharge-charge by SEM (Fig. 1). As the discharge proceeds (i-iii), an increasing number of cubic $NaO_2$ crystals is precipitated at the rGO surface. As they become larger, the morphology of the crystals become less perfect, with clear stepped surfaces and misaligned crystal planes. On charge, $NaO_2$ crystals seem to dissolve from the surface, preferentially at the high miller index corner sites and the terrace edges, suggesting lower dissolution energies for higher miller index planes. As charging proceeds, the dissolving corners merge together and the crystal shrinks eventually to a small hemispherical particle before complete disappearance. The fact that $NaO_2$ decomposition occurs at the crystal surface away from the rGO-$NaO_2$ interface (iv-viii) also supports a dominant dissolution process during charge rather than a direct electrochemical decomposition via charge transport through the poorly conductive $NaO_2$ crystal bulk, as the latter would be expected to be more facile at the rGO-$NaO_2$ interface. Nevertheless, the latter mechanism can indeed occur at some local regions, as evidenced by interfacial decomposition reactions for some particles (Fig. S4). To operate via the solution mechanism, bare rGO surfaces and water molecules are needed. After a deep discharge, however, rGO surfaces are heavily covered by $NaO_2$ crystals (Fig. S1B), so that interfacial decomposition of $NaO_2$ may become favoured over the solution process at some local regions.

Although PXRD and SEM results support a reversible formation and decomposition of $NaO_2$ primarily via a solution mechanism, the cycling ability of the battery is poor.

Even on limiting the capacity to 0.5 mAh per discharge (1 mAh/cm$^2$), the cell rapidly fails after 18 cycles (Fig. 1(b)), the coulombic efficiency being around 92%. This means up to 8% of the NaO$_2$ phase formed in each cycle ends up blocking the cell or forming side reaction products. To confirm this, SEM measurement was performed on an rGO electrode after 1$^{st}$ charge (Fig. S5). Some residual NaO$_2$ was indeed observed and it seems that NaO$_2$ was more thoroughly removed in the central regions than the peripheral regions of the electrode. In addition, compared with the pristine rGO surface (Fig. S6(a-b)), the charged rGO electrode of the 1$^{st}$ cycle also contains large particles(c-d) that look different from NaO$_2$. After 20 cycles, more particles of this type up to 100 μm in size were found clogging the pores within the rGO electrode, (Fig. S6(e)); at a smaller scale, the rGO surface that was initially smooth and clean (b) is now covered by numerous small particles (f-g) and some thin films (g-h) due to side reactions. Given that the rGO electrode still looks porous and the electrolyte in the cell had not dried out at the end of cycling, we conclude that the passivation of the rGO surface by side reaction products is more likely to be the main cause of cell failure. Next we investigated the nature of the side reactions occurring in the Na-O$_2$ battery using solid-state NMR (ssNMR) and Raman spectroscopy.

Six Na-O$_2$ cells were cycled and $^{23}$Na and $^1$H ssNMR measurements of the rGO electrodes were performed on cells after the 1$^{st}$, 5$^{th}$ and 8$^{th}$ discharge and charge cycles. $^{23}$Na ssNMR spectra (Fig. 2(a)) show that the discharged sample has a major resonance at -27 ppm attributed to NaO$_2$ with two clear shoulders at around -16 and -5 ppm, the resonance at -27 ppm is attributed to $NaO_2$, which reversibly appears and disappears in the corresponding discharge and charge spectra and is in good agreement with the previously reported spectrum of the room-temperature $NaO_2$ (I) phase.[20] The other two shoulders are more difficult to assign due to their featureless lineshapes, but since there are no significant intensity changes between discharge and charge, this suggests that they can be ascribed to a mixture of various side reaction products. With increasing number of cycles, some residual $NaO_2$ (-27 ppm) clearly accumulates, together with a broad resonance spanning -40 to 20 ppm. These observations are consistent with the XRD (Fig. 1(c)) and SEM results (Fig. S5-6). The irreversibility is more clearly revealed by the corresponding $^1H$ NMR spectra (Fig. 2(b)). For both discharge and charge, resonance intensities between -10 and 10 ppm continue to increase with the number of cycles, exhibiting 4 major components at around -2.8, 1.8, 3.3 and 8.5 ppm. To aid the interpretation of these peaks, a library of Na reference compounds were measured (Fig. S7) and the above 4 resonances are assigned, in sequence, to sodium hydroxide, acetate, methoxide (possibly with other ether fragments) and formate. These Na species do not seem to be removed during charge. Of note, in order to further resolve sodium-containing species, a 2-dimensional (2D) $^1H$-$^{23}Na$ heteronuclear correlation (HETCOR) experiment was acquired with a cell finishing at the end of charge in the 8th cycle. However, we only observed sodium hydroxide (Fig. S10a) and this may be due to the low concentration of other species in the cycled sample. A highly shifted $^1H$ resonance at 29 ppm was also observed and the corresponding species seems to be reversibly formed and removed during cycling. The $^1$H $T_1$ relaxation time of 4.0 ± 0.4 ms is obtained from the 8$^{th}$ discharged sample for this resonance and this is two orders of magnitude shorter than a value for the formate (0.41 ± 0.02 s), indicating that this species is nearby a paramagnetic centre. On this basis we tentatively ascribe this resonance to a proton species in a close proximity to the NaO$_2$ phase, either on its surface, or replacing a Na$^+$ in the bulk, resulting in a large paramagnetic shift (and consistent with change in cell parameters of the NaO$_2$ phase seen on cycling).

It is known[21,22] that carbonate is another common decomposition product in ether-based Li-O$_2$ batteries. To examine this possibility in Na-O$_2$ cells, we conducted $^{13}$C ssNMR measurements (Fig. 2(c)) on the rGO electrode from the cell finishing at the end of charge in the 8$^{th}$ cycle. Compared with the $^{13}$C spectrum of pristine rGO, the cycled electrode shows additional resonances between 0-30 ppm (*sp$^3$* carbon) and an intense resonance at around 160-190 ppm, the latter being a mixture of sodium formate, carbonate and acetate.[23] The Raman measurements on a multiple cycled electrode further confirm that sodium carbonate, hydroxide and acetate are three major components of the side reaction product (Fig. S8), consistent with the ssNMR results.

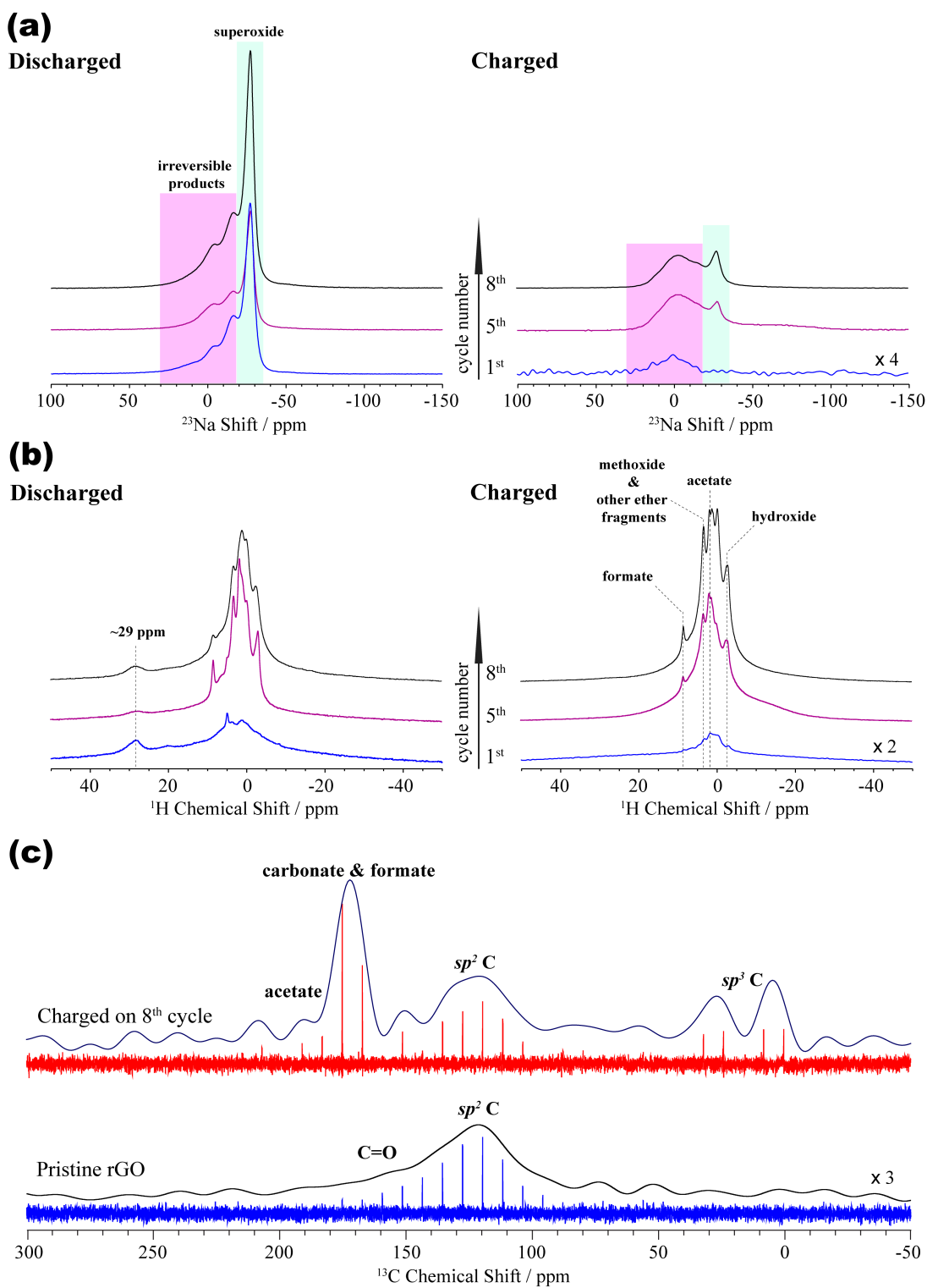

**Figure 2.** $^{23}$Na (a), $^{1}$H (b), and $^{13}$C (c) ssNMR spectra of cycled electrodes, acquired at 11.7 T, with a MAS frequency of 60 kHz. The major discharge product, $NaO_2$ and a spectral region containing of irreversible side reaction products are highlighted on $^{23}$Na spectra. The irreversible products: hydroxide, acetate, methoxide, and formate (-2.8, 1.8, 3.3 and 8.5 ppm, respectively) are assigned on the $^{1}$H spectra.

The product formed on discharge needs to be chemically stable in the electrolyte, this property directly determining the long-term stability of the cell. The chemical reactivity of the $NaO_2$ and $Na_2O_2$ phases in DME was examined after soaking them with DME for 15 days. Fig. 3 compares the $^{23}$Na (a) and $^1$H (b) ssNMR spectra of pristine $NaO_2$ (electrochemically formed) and $Na_2O_2$ to the spectra obtained from the corresponding soaked samples. It is clear that additional sodium species are formed after soaking. For $NaO_2$, two additional $^{23}$Na NMR resonances at around 5 and 12 ppm appear, which largely resemble the discontinuities of the $2^{nd}$-order quadrupolar lineshape of sodium peroxide and overlap with the other resonances of sodium carbonate, acetate and formate (see Fig. S7 and S11). The corresponding $^1$H NMR spectrum shows similar resonances to those prior to mixing but the resonances are less well resolved. The intensity of the 2-10 ppm region is increased, also suggesting the formation of acetate and formate. SEM reveals that the side reaction products formed a thin film covering the $NaO_2$ crystals (Fig. S9). Such a surface layer will only partially suppress further side reactions with $NaO_2$, because the repeated formation/decomposition of $NaO_2$ during cycling will continue to rupture the layer and expose fresh $NaO_2$ surfaces. The soaked $Na_2O_2$ sample also exhibits side reactions, NaOH being more prominent than the formate, acetate and methoxide species (Fig. 3a and b). The dominance of NaOH in the decomposition products is further confirmed by a 2D $^1$H-$^{23}$Na HETCOR experiment (Fig. S10b).

From the above microscopic and spectroscopic investigations, it is clear that the sodium superoxide and peroxide phases are not stable even for 15 days in an ether electrolyte (DME): they cause electrolyte decomposition via nucleophilic attack (dominated by superoxide) or proton extraction (dominated by peroxide), similar to that observed for $Li_2O_2$ and $KO_2$.[21,24-26] Furthermore, the solution mechanism enabled by water can lead to an increased concentration[27] and lifetime of the chemically aggressive superoxide anion in the non-aqueous electrolyte and it is expected to accelerate electrolyte decomposition. Using higher donor number solvents that stabilize the superoxide anion (e.g., dimethyl sulfoxide) has been reported to cause the same problem.[24,28] On the other hand, the low overpotential during charge in $Na-O_2$ cells can help minimize the interfacial carbonate formation between the discharge product and carbon electrode,[29] and thus somewhat slow down cell failure.

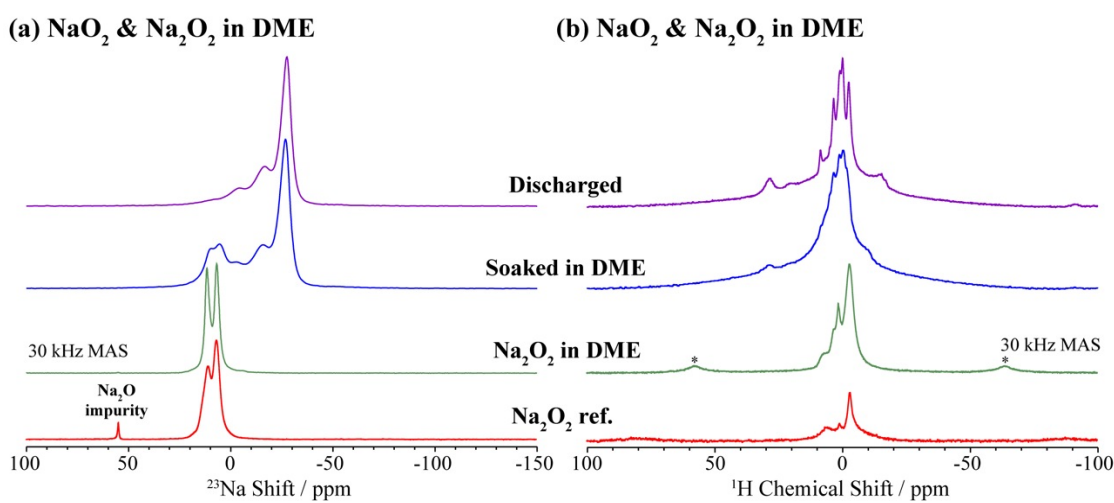

**Figure 3**. Demonstrating chemical reactivity of electrochemically-prepared $NaO_2$ and $Na_2O_2$ by (a) $^{23}Na$ and (b) $^{1}H$ ssNMR spectroscopy. All NMR spectra were acquired at 11.7 T, with a MAS rate of 60 ($^{23}Na$ (a) and $^{1}H$ (b), otherwise stated) and 30 ($^{1}H$ (c)) kHz.

To realize a long-life superoxide based cell, it is worth considering the following aspects. An electrolyte needs to be found or synthesized that is stable against the superoxide radicals; the stability of an electrolyte can be tested by exposing it to a superoxide phase for a long period. The chemical instability of superoxide species against the electrolyte is probably the most fundamental challenge for metal-oxygen batteries based on $LiO_2$, $NaO_2$ or $KO_2$. When a solution mechanism dominates the formation and decomposition of the superoxide phase, factors controlling the concentration, diffusion of the participating species and the spatial distribution of the discharge products need to be considered to enable a homogeneous reaction across the electrode surface and thickness. However, the higher solubility of superoxide species in the electrolyte means they are able to diffuse further towards to the anode and cause detrimental effects. The deposition and accumulation of superoxide on the separator may also gradually impede ion transport; the electrically isolated discharge product is also partially responsible for the inability to fully recharge the $Na-O_2$ cell. These fundamental challenges need to be addressed before realizing an efficient and long-lived $Na-O_2$ battery.

**ACKNOWLEDGEMENTS**

The authors thank EPSRC (TL, GK and CPG) and Innovate UK (TL) for research funding. This project has received funding from the European Unions's Horizon 2020 research and innovation programme under grant agreement No. 696656 – GrapheneCore1 (GK and CPG). Dr. Elizabeth Castillo-Martínez and Dr. Zigeng Liu are thanked for useful discussions. GK thanks Mr. Jeongjae Lee for preliminary DFT calculations and related discussions.

**AUTHOR INFORMATION**

Corresponding Author

C.P.G.: Email: cpg27@cam.ac.uk

Notes

The authors declare no completing financial interest.

**ASSOCIATED CONTENT**

Supporting Information is free of charge at the ACS publication website (see attached file).

**REFERENCE**

(1) Bruce, P. G.; Freunberger, S. A.; Hardwick, L. J.; Tarascon, J.-M., Li-O$_2$ and Li-S Batteries with High Energy Storage. *Nat. Mater.* **2012**, *11,* 19.

(2) Hartmann, P.; Bender, C. L.; Vracar, M.; Durr, A. K.; Garsuch, A.; Janek, J.; Adelhelm, P., A Rechargeable Room-temperature Sodium Superoxide (NaO$_2$) Battery. *Nat. Mat.* **2013**, *12,* 228-232.

(3) Ren, X.; Wu, Y., A Low-overpotential Potassium-oxygen Battery Based on Potassium Superoxide. *J. Am. Chem. Soc.* **2013**, *135,* 2923-2926.

(4) Lu, J.; Lee, Y. J.; Luo, X.; Lau, K. C.; Asadi, M.; Wang, H. H.; Brombosz, S.; Wen, J.; Zhai, D.; Chen, Z.; *et al.* A Lithium-oxygen Battery Based on Lithium Superoxide, *Nature* **2016**, *529,* 377-382.

(5) Abraham, K. M.; Jiang, Z. A Polymer Electrolyte Based Rechargeable Lithium/oxygen Battery. *J. Electrochem. Soc.* **1996**, *143,* 1-5.

(6) Ogasawara, T.; Debart, A.; Holzapfel, M.; Novak, P.; Bruce, P. G. Rechargeable Li$_2$O$_2$

Electrode for Lithium Batteries. *J. Am. Chem. Soc*. **2006**, *128,* 1390-1393.

(7) Liu, W.; Sun, Q.; Yang, Y.; Xie, J.; Fu, Z. An Enhanced Electrochemical Performance of a Sodium-air Battery with Graphene Nanosheets as Air Electrode Catalysts. *Chem. Commun*. **2013**, *49,* 1951-1953.

(8) Liu, T. Leskes, M.; Yu, W.; Moore, A.; Zhou, L.; Bayley, P.; Kim, G.; Grey, C. P. Cycling Li-$O_2$ Batteries via LiOH Formation and Decomposition. *Science*, **2015**, *350,* 530-533.

(9) Li, F.; Wu, S.; Li, D.; Zhang, T.; He, P.; Yamada, A.; Zhou, H. The Water Catalysis at Oxygen Cathodes of Lithium-oxygen Cells. *Nat. Commun*. **2015**, *6,* 7843.

(10) Xia, C.; Black, R.; Fernandes, R.; Adams, B.; Nazar, L. F. The Critical Role of Phase-transfer Catalysis in Aprotic Sodium Oxygen Batteries. *Nat. Chem*. **2015**, *7,* 496-501.

(11) Beder, C. L.; Hartmann, P.; Vracar, M.; Adelhelm, P.; Janek, J. On the Thermodynamics, the Role of the Carbon Cathode, and the Cycle Life of the Sodium Superoxide ($NaO_2$) Battery. *Adv. Energy, Mat.* **2014**, *4,* 1301863.

(12) Kim, J.; Park, H.; Lee, B.; Seong, W. M.; Lim, H.-D.; Bae, Y.; Kim, H.; Kim, W. K.; Ryu, K. H.; Kang, K. Dissolution and Ionization of Sodium Superoxide in Sodium-oxygen Batteries. *Nat. Commun*. **2016**, *7,* 10670.

(13) Jian, Z.; Chen, Y.; Li, F.; Zhang, T.; Liu, C.; Zhou, H. High Capacity Na-$O_2$ Batteries with Carbon Nanotube Paper as Binder-free Air Cathode. *J. Power Sources* **2014**, *251,* 466.

(14) Yadegari, H.; Li, Y.; Banis, M. N.; Li, X.; Wang, B.; Sun, Q.; Li, R.; Sham, T. K.; Cui, X.; Sun, X. On Rechargeability and Reaction Kinetics of Sodium-air Batteries. *Energy Environ. Sci*. **2014**, *7,* 3747.

(15) Zhao, N.; Li, C.; Guo, X. Long-life Na-$O_2$ Batteries with High Energy Efficiency Enabled by Electrochemically Splitting $NaO_2$ at a Low Overpotential. *Phys. Chem. Chem. Phys*. **2014**, *16,* 15646.

(16) Pinedo, R.; Weber, D. A.; Bergner, B.; Schroder, D.; Adelhelm, P.; Janek, J. Insights into the Chemical Nature and Formation Mechanisms of Discharge Products in Na-$O_2$ Batteries by Means of Operando X-ray Diffraction. *J. Phys. Chem. C* **2016**, *120,* 8472.

(17) Kwabi, D. G.; Tulodziecki, M.; Pour, N.; Itkis, D. M.; Thompson, C. V.; Shao-Horn, Y. Controlling Solution-Mediated Reaction Mechanisms of Oxygen Reduction Using Potential and Solvent for Aprotic Lithium−Oxygen Batteries. *J. Phys. Chem. Lett*. **2015**, *6,* 2636-2643.

(18) Hartmann, P.; Heinemann, M.; Bender, C. L.; Graf, K.; Baumann, R. P.; Adelhelm, P.; Heiliger, C.; Janek, J. Discharge and Charge Reaction Paths in Sodium-Oxygen Batteries: Does $NaO_2$ Form by Direct Electrochemical Growth or by Precipitation from Solution? *J. Phys. Chem. C* **2015**, *119,* 22778.

(19) Lee, B.; Kim, J.; Yoon, G.; Lim, H. D.; Choi, I. S.; Kang, K. Theoretical Evidence for Low Charging Overpotentials of Superoxide Discharge Products in Metal-Oxygen Batteries, *Chem. Mater*. **2015**, *27,* 8406.

(20) Krawietz, T. R.; Murray, D. K.; Haw, J. F. Alkali Metal Oxides, Peroxides, and Superoxides: A Multinuclear MAS NMR Study. *J. Phys. Chem*. A, **1998**, *102,* 8779-8785.

(21) Freunberger, S. A.; Chen, Y.; Drewett, N. E.; Hardwick, L. J.; Bardé, F.; Bruce, P. G. The Lithium−Oxygen Battery with Ether-Based Electrolytes. *Angew. Chem. Int. Ed.* **2011**, *50,* 8609.

(22) Leskes, M.; Moore, A. J.; Goward, G. R.; Grey, C. P. Monitoring the Electrochemical Processes in the Lithium−Air Battery by Solid State NMR Spectroscopy. *J. Phys. Chem. C* **2013**,*117,* 26929.

(23) Gottlieb, H. E.; Kotlyar, V.; Nudelman, A. NMR Chemical Shifts of Common Laboratory Solvents as Trace Impurities. *J. Organ. Chem*. **1997**, *62,* 7512-7515.

(24) Kwabi, D. G.; Batcho, T. P.; Amannchukwu, C. V.; Ortiz-Vitoriano, N.; Hammond, P.; Thompson, C. V.; Shao-Horn, Y. Chemical Instability of Dimethyl Sulfoxide in Lithium−Air Batteries. *J. Phys. Chem. Lett*. **2014**, *5,* 2850-2856.

(25) Sharon, D.; Hirshberg, D.; Afri, M.; Garsuch, A.; Frimer, A. A.; Aurbach, D. Lithium-oxygen Electrochemistry in Non-aqueous Solutions. *Israel J. Chem*. **2015**, *55,* 508-520.

(26) Black, R.; Oh, S. Y.; Lee, J. H.; Yim, T.; Adams, B.; Nazar, L. F. Screening for Superoxide Reactivity in Li-O$_2$ Batteries: Effect on Li$_2$O$_2$/LiOH Crystallization. *J. Am. Chem. Soc*. **2012**, *134,* 2902-2905.

(27) Xia, C.; Fernandes, R.; Cho, F. H.; Sudhakar, N.; Buonacorsi, B.; Walker, S.; Xu, M.; Baugh, J.; Nazar, L. F. Direct Evidence of Solution-Mediated Superoxide Transport and Organic Radical Formation in Sodium-Oxygen Batteries. *J. Am. Chem. Soc*. **2016**, *138,* 11219-11226.

(28) Sharon, D.; Afri, M.; Noked, M.; Garsuch, A.; Frimer, A. A.; Aurbach, D. Oxidation of Dimethyl Sulfoxide Solutions by Electrochemical Reduction of Oxygen. *J. Phys. Chem. Lett*. **2013**, *4,* 3115-3119.

(29) Thotiyl, M. M. O.; Freunberger, S. A.; Peng, Z.; Bruce, P. G. The Carbon Electrode in Non-aqueous Li-O$_2$ Cells. *J. Am. Chem. Soc.* **2013**, *135,* 494.